\documentclass[conference]{IEEEtran}
\usepackage{blindtext, graphicx}
\graphicspath{ {Images/} }

\usepackage[english]{babel}
\usepackage{comment}
\usepackage{csquotes}

\usepackage[
backend=biber,
style=numeric,
sorting=ynt
]{biblatex}
\usepackage{biblatex}
\usepackage{ragged2e}

\ifCLASSINFOpdf
  
\else
  
\fi

\usepackage{balance}
\usepackage[T1]{fontenc}
\usepackage{mathptmx}
\usepackage{subcaption} 
\renewcommand{\figurename}{Fig.}
\addto\captionsenglish{\renewcommand{\figurename}{Fig.}}
\begin{document}
%
\title{\Huge{Vision Transformer based COVID-19 Detection using Chest X-rays}}

\author{\IEEEauthorblockN{\small{Koushik Sivarama Krishnan}}
\IEEEauthorblockA{
\small{Email: koushik.nov01@gmail.com}}
\and
\IEEEauthorblockN{\small{Karthik Sivarama Krishnan}}
\IEEEauthorblockA{
\small{Email: ks7585@rit.edu}}}

\maketitle

\small{\emph{\textbf{Abstract}} - \textbf{COVID-19 is a global pandemic, and detecting them is a momentous task for medical professionals today due to its rapid mutations. Current methods of examining chest X-rays and CT scan requires profound knowledge and are time consuming, which suggests that it shrinks the precious time of medical practitioners when people's lives are at stake. This study tries to assist this process by achieving state-of-the-art performance in classifying chest X-rays by fine-tuning Vision Transformer(ViT). The proposed approach uses pretrained models, fine-tuned for detecting the presence of COVID-19 disease on chest X-rays. This approach achieves an accuracy score of 97.61\%, precision score of 95.34\%, recall score of 93.84\% and, f1-score of 94.58\%. This result signifies the performance of transformer-based models on chest X-ray.}}


\IEEEpeerreviewmaketitle

\section{\normalsize{Introduction}}
    
    The term Global Pandemic applies to an epidemic of an infectious disease that is spread all across the globe. The world has not seen a global pandemic after the Spanish flu up until the year 2019. In late 2019, a new strain of corona virus disease was found in Wuhan district of Hubei Province in China. Since this strain of coronavirus disease was not previously identified in human beings, World Health Organization named this strain as "Novel" coronavirus disease. The Novel Coronavirus Disease of 2019 was officially given an abbreviated name of COVID-19 by the World Health Organization,  where the acronym CO stands for \textbf{Co}rona, VI stands for \textbf{Vi}rus, D stands for \textbf{D}isease followed by the year 2019 represented as \textbf{19}.    \\\\
    This disease abruptly spread all across the globe affecting millions of people and created global tension across all countries.  Covid-19 is caused by the Severe Acute Respiratory Syndrome Coronavirus 2 (SARS-CoV-2) \cite{walls2020structure}. This virus reside within the same clan of Severe Acute Respiratory Syndrome (SARS) and Middle East Respiratory Syndrome (MERS).The commonly identified symptoms for  this disease include coughing, shortness of breath, fever, pneumonia, and severe respiratory distress. The symptomatology of this disease is so diverse and are usually identified within day one to day fourteen after the person is being exposed to this contagious virus. \\
    
    After the initial identification of the presence of covid 19 disease in China, the cases started expanding exponentially, affecting millions of people. This sudden burst of cases, forced all countries to take severe measures to keep the outbreak in control. Figure 1 shows the chart  of new cases from January 2020 to July 2021 published in World Health Organization website.
    \begin{figure}[h]
    \includegraphics[width = 8cm, height = 6cm]{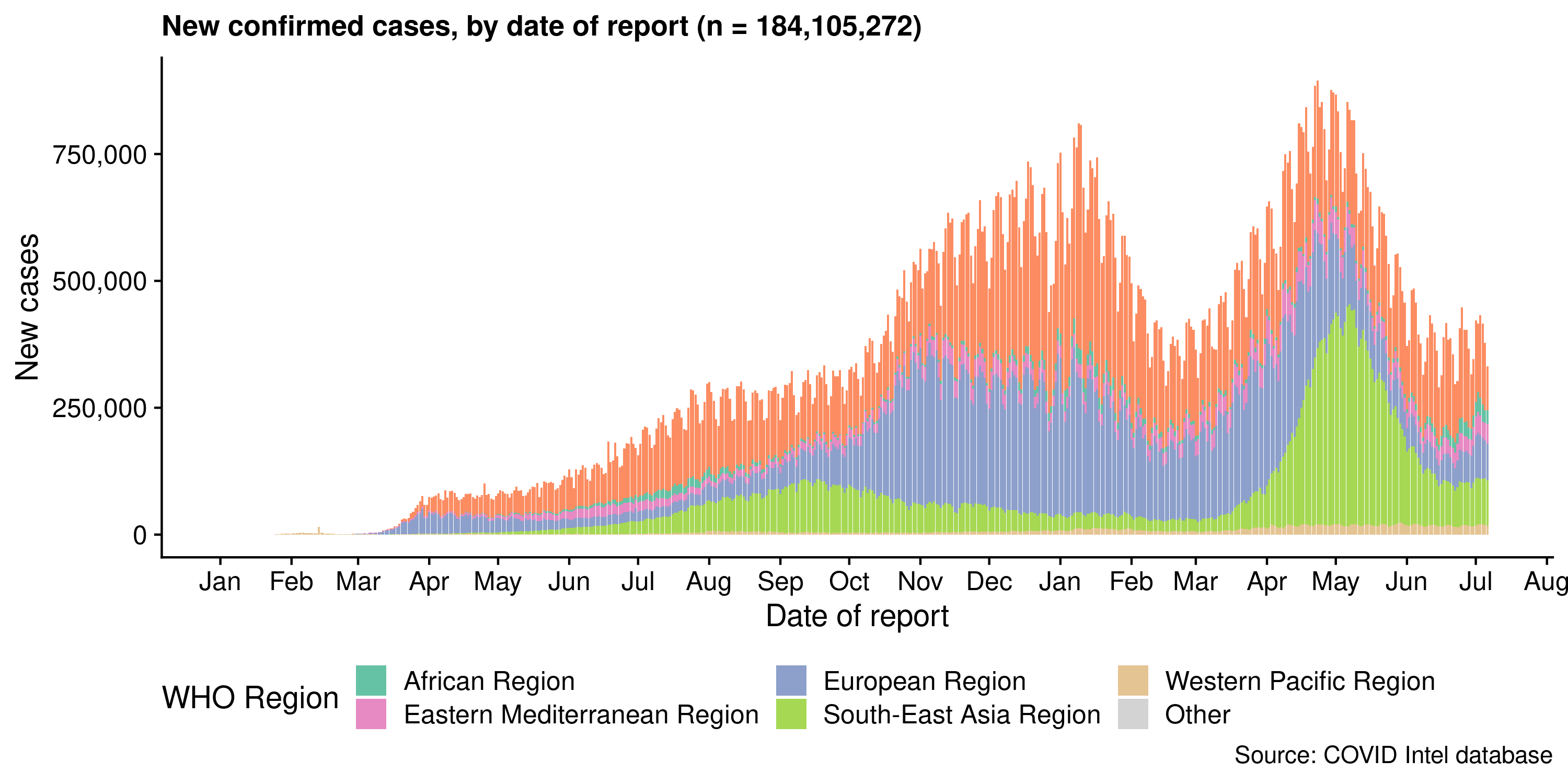}
    \caption{\footnotesize{COVID-19 cases worldwide published by WHO \cite{covid19explorer}}}
    \label{fig:subim1}
    \end{figure}\\\\
    With an exponential increase in covid 19 cases,there is a high demand for rapid testing. Baseline gold standard approach used to test this coronavirus disease 2019 is reverse transcription polymerase chain reaction (RT-PCR). The results for this test takes somewhere between 48 to 72 hours of wait time for people to know if they have been affected by this contagious virus or not.  Other testing methods include Antigen testing for coronavirus disease. The results of this test usually takes around 15 minutes to 30 minutes of wait time for people to know if they have been exposed to the contagious virus or not. Typically, patients experiencing symptoms of this coronavirus disease 2019 are asked to take Chest CT scans or X-ray scans. Then a radiologist expert examines the scans to confirm if there is a presence of this disease in the chest area. This process is time consuming and is hardly inefficient considering the rise in cases every minute. Rapid and rigorous testing is required in-order to identify as much as cases possible and prevent them from spreading this virus to others. This calls for the need of artificial intelligence to help identify covid-19 patients with minimum time consumption and maximum efficiency.  \\
    
    Artificial Intelligence and Machine learning/Deep learning are currently the state of the art in predicting results in almost all fields. The AI based systems are showing massive results in healthcare domain and this would be helpful in mitigating the time delay for covid-19 virus identification. The vision transformer is the latest state-of-the-art deep learning architecture as it achieves similar performance in benchmark datasets compared to CNN-based models and sometimes even outperforms them in some tasks.

\section{\normalsize{Related Work}}
    Deep Learning has become a great area of interest for researchers in recent days. Many have successfully applied deep learning / machine learning techniques \cite{8250154} for medical classification and obtained promising results. The transfer learning-based approaches have shown remarkable performance boosts in almost all Computer Vision and Natural Language Processing tasks. This approach saves a lot of computational time and can be trained with a fraction of the data.

\normalsize{\emph{\subsection{Transfer Learning}}}
    Transfer Learning has become a common practice in almost all Natural Language Processing and Computer Vision tasks, given the enormous computational power required to train them. This approach is a two-step process. At first, you train the model on large volumes of benchmark datasets. Then you fine-tune them by freezing the initial layers and replacing the last layer based on the task at hand, thus using the previously learnt features for representation. This approach requires a fraction of the data required to train a model from scratch and can be trained in a shorter duration, with less computational resources.
    
\normalsize{\emph{\subsection{Transformers}}}
    The Transformer architecture\cite{vaswani2017attention} was initially proposed in 2017 for neural machine translation. This opened up a wide range of new possibilities in the domain of Natural Language Processing. Since then, several variants and advancements were proposed in transformer-based architectures and all of them are generally pretrained on a large corpus of text data. This architecture proved that combining self-attention with linear layers outperformed the traditional sequence-to-sequence LSTM-RNN based approaches in  Neural Machine Translation and other Natural Language Processing tasks. 

\normalsize{\emph{\subsection{Vision Transformers}}}
    Since the introduction of transformer architecture\cite{vaswani2017attention}, many researchers have tried to apply them to Computer Vision tasks. Many even added CNN before self-attention to extract features from images, but most of these approaches could not be applied to real-world Computer Vision problems. The successful utilization of transformer architecture for images was proposed by Dosovitskiy et al. \cite{dosovitskiy2021image}. They managed to get higher accuracy with less computational time for training. This approach also has a much less image-specific bias when compared to CNN-based models. In this approach, they split the images into small patches and stacked them linearly to pass them as input to the transformer model.

\normalsize{\emph{\subsection{Transfer learning based approaches}}} 
    Talha Anwar and others. \cite{9318212} used the pretrained EfficientNet family of architectures to detect COVID-19 in CT scans and managed to get an accuracy score of 89.7\%, an F1 score of 89.6\%, and an AUC score of 89.5\%. \\\\
    The authors, Rahaman and others. \cite{Rahaman2020} have developed a CAD system for detecting COVID-19 X-rays from others by fine-tuning the pretrained VGG19 model. They managed to obtain a precision score of 97.5\% and a recall score of 82.0\%. This study experimented with various pretrained CNN-based models and found VGG19 outperforming other latest models.\\\\
    Sethy et al.\cite{sethy2020detection} suggested a classification model that combined both pretrained ResNet50 and SVM. Here, the author used the Convolutional Neural Network to extract feature from images. This extraced features is then passed into SVM. They showed that this approach is superior to using just a CNN-based network. They achieved an FPR of 95.38\%, F1 score of 95.52\%, MCC score of 91.41\%, and Kappa score of 90.76\% for detecting COVID-19 in X-ray images. The authors of \cite{8753848} also used convolutional neural networks as a feature extraction technique.

\section{\normalsize{Methodology}}
\normalsize{\emph{\subsection{Dataset}}}
    To test how good our proposed method is, we used chest X-rays of COVID19 patients, healthy patients, patients with opaque lungs, and patients with viral pneumonia. We collected images from two publicly available data sets. \\\\ 
    COVID19 X-ray database dataset \cite{9144185} \cite{RAHMAN2021104319}. This was created by a team of researchers from various universities, in cooperation with doctors from Malaysia. They examined various X-rays and classified them into COVID-19, Viral Pneumonia, Lung Opacity and Healthy chest X-rays. \\\\
    The COVID19 Pneumonia Normal Chest X-ray PA Dataset from Kaggle was used as the test set.\\\\
    COVID-19 image from the COVID19 X-ray database\cite{9144185} \cite{RAHMAN2021104319} was originally around 3500. We upsampled it to 7,000 by using various image augmentation techniques. We then took equal samples of images from Viral Pneumonia, Lung Opacity and Healthy chest X-rays for the Non-COVID class. 
\begin{figure}[ht]
\begin{center}
\includegraphics[width = 4cm, height = 4cm]{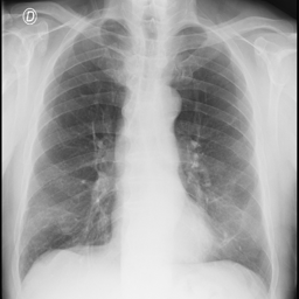}
\caption{\footnotesize{(a) COVID}}
\end{center}

\label{fig:subim2}

\begin{center}
\includegraphics[width = 4cm, height = 4cm]{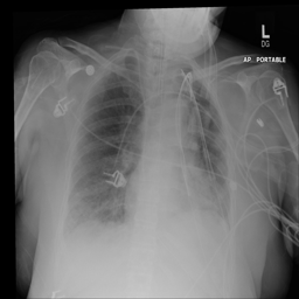}

\caption{\footnotesize{(b) Lung Opacity}}
\label{fig:subim3}
\end{center}

\begin{center}
\includegraphics[width = 4cm, height = 4cm]{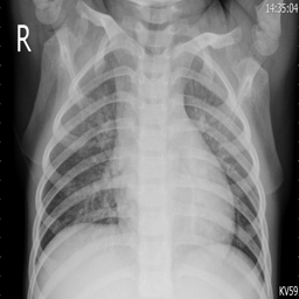}
\caption{\footnotesize{(c) Viral Pneumonia}}
\label{fig:subim4}
\end{center}

\begin{center}
\includegraphics[width = 4cm, height = 4cm]{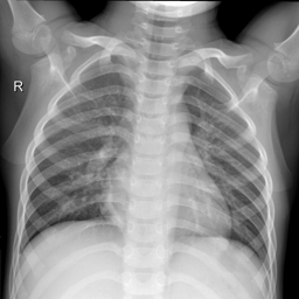}
\caption{\footnotesize{(d) Normal}}
\label{fig:subim5}
\end{center}


\end{figure}
\centerline{\footnotesize{I Dataset Split}}
    
   \noindent \begin{tabular}{|p{18mm}|p{18mm}|p{20mm}|p{18mm}|}
    
    \hline
    Dataset & COVID & NON-COVID & TOTAL\\
    \hline
    Train    & 6880 & 6980 & 13,860\\
    Validation  & 350 & 369 & 719\\
    Test  & 2313 & 2313 & 4626\\
    Total   & 9443 & 9662 & 19,105\\
    \hline
    \end{tabular}\\\\

\normalsize{\emph{\subsection{Dataset Preprocessing}}}
    The COVID-19 images are up-sampled using image augmentation techniques with the help of albumentations \cite{info11020125} library. The chest X-ray is flipped vertically or horizontally, then randomly rotated to the limit of 270 degrees  with constant edges, and the brightness and contrast are arbitrarily adjusted to the limit of 0.4. \\\\ 
    chest X-rays are of different sizes from 447 × 530 to 4200 × 3290 pixels. Therefore, we made the target size of images as  224 x 244 pixels. As the models were pretrained on RGB images, we created fake RGB images by stacking the channel over itself. Contrast Limited Adaptive Histogram Equalization (CLAHE) \cite{info11020125} is used as an image enhancement method. CLAHE is a revision of Adaptive Histogram Equalization (AHE) that avoids excessive contrast enhancement in the image.\\

\section{\normalsize{Experiments and Analysis}}
\normalsize{\emph{\subsection{Experimental Setup}}}
    In this research, we have used DenseNet\cite{huang2018densely}, InceptionV3 \cite{szegedy2015rethinking}, WideResNet101\cite{DBLP:journals/corr/ZagoruykoK16} and, Vision Transformer(ViT-B/32)\cite{dosovitskiy2021image} models in the transfer learning process, where the model is pre-trained on the ImageNet \cite{5206848} dataset. The bottom layer of the architecture is frozen, as it only extracts general features. We replaced the top layer of the model so that it returns dataset-specific output from the linear layer to match our dataset.\\\\
    In ViT-B/32\cite{dosovitskiy2021image} pretrained model, the images are split into 32x32 patches and passed into an embedding layer to get the patch embeddings. As the transformers are permutationally invariant, 1-D position embeddings extracted from the images are added to the patch embeddings. This resultant embedding vector is then fed into the encoder part of ViT\cite{dosovitskiy2021image}. The encoder of ViT\cite{dosovitskiy2021image} consists of alternating Layer Normalization, Multiheaded Self-Attention and, MLP\cite{tolstikhin2021mixer} blocks. Skip connections are applied after every block.
    
    \begin{figure}[ht]
    \includegraphics[width = 9cm, height = 10.2cm]{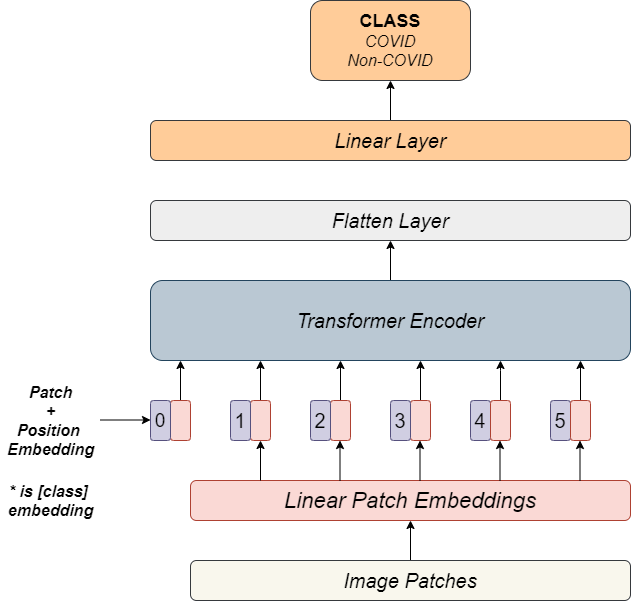}
    \caption{\footnotesize{Model Architecture}}
    \label{fig:subim6}
    \end{figure}
    The logits from the linear layer pass through a sigmoid activation function, which returns the predicted probabilities. The target size of the image was set as 224 x 224 x 3 for all models, with an initial learning rate of 1e-4. We used Adam \cite{Adam} optimizer for CNN based models and RectifiedAdam \cite{liu2020variance} optimizer for ViT-B/32\cite{dosovitskiy2021image} with the "ReduceLROnPlateau" and "EarlyStopping" schedulers. The learning rate and optimizer were found using hyperparameter optimization framework Optuna \cite{akiba2019optuna}.\\\\ 
    We used Optuna\cite{akiba2019optuna} for 50 trials, using various optimizers and learning rates from 1e-3 to 1e-6 and found Adam optimizer works well for all pre-trained CNN models plus 1e-4 as the best initial learning rate.  We used the RectifiedAdam\cite{liu2020variance} optimizer as it showed considerable performance improvement over vanilla Adam\cite{Adam} optimizer for the ViT\cite{dosovitskiy2021image} pre-trained model. ReduceLROnPlateau scheduler monitors the validation accuracy and reduces the learning rate by a factor of 0.2 if it does not improve for more than three epochs. EarlyStopping scheduler monitors validation accuracy and stops training if the monitored metric does not improve.\\\\
    InceptionV3\cite{szegedy2015rethinking}, DenseNet\cite{huang2018densely} and, WideResNet101\cite{DBLP:journals/corr/ZagoruykoK16} were trained on Google Colaboratory with a batch size of 32 for 25 epochs. ViT-B/32\cite{dosovitskiy2021image} model was trained on NVIDIA RTX 2060 mobile GPU with a batch size of 16 and was early stopped at an epoch of 10.

\normalsize{\emph{\subsection{Evaluation Criteria}}}
    In this study, we used Accuracy, Precision, Recall, and f1-score as scoring metrics. The accuracy gives a basic idea of the model's performance and is simply a ratio of the correctly predicted observation to the overall observations. The ratio of the correctly predicted positive output and the entire predicted positive output is known as precision score. The recall score is the ratio of the correctly predicted positive output and the total output of the actual positive observation. The weighted average of precision score and recall score is referred to as the F1 score, where the maximum value is 1 and the minimum value is 0.
    
    \begin{equation}
    Accuracy  = \frac{TP+TN}{TP+FP+FN+TN}
    \end{equation}
    \begin{equation}
        Precision = \frac{TP}{TP+FP}
    \end{equation}
    \begin{equation}
        Recall = \frac{TP}{TP+FN}
    \end{equation}
    \begin{equation}
        F1-score = 2 * \frac{Recall * Precision}{Recall + Precision}
    \end{equation}

\section{\normalsize{Results}}
    To derive conclusions, we compared the accuracy, precision, recall and F1-score of each model.\\\\
    \centerline{\footnotesize{II Performance Analysis Table}}
    
   \noindent \begin{tabular}{|p{21mm}|p{15mm}|p{12mm}|p{10mm}|p{11mm}|}
    
    \hline
    Model & A.Precision & A.Recall & A.F1 & ACC(\%)\\
    \hline
    InceptionV3    & 0.875 & 0.870 & 0.872 & 0.88\\
    DenseNet       & 0.910 & 0.922 & 0.915 & 0.92\\
    WideResNet101  & 0.865 & 0.855 & 0.845 & 0.85\\
    \textbf{ViT-B/32}   & \textbf{0.953} & \textbf{0.938} & \textbf{0.946} & \textbf{0.976}\\
    \hline
    \end{tabular}\\\\\\
    The above table illustrates the performance of different deep learning models for detection of COVID-19 from chest X-rays. The test accuracy gives us a general idea about the model's real world performance.\\\\
    It can be observed from the above table that the highest avg. precision, avg. recall, avg. F1-score and, test accuracy of 95.3\%, 93.8\%, 94.6\%, 97.6\% respectively was achieved by ViT-B/32\cite{dosovitskiy2021image} model, followed by DenseNet\cite{huang2018densely} with value of 91.0\%, 92.2\%, 91.5\% and, 92\% respectively. The WideResNet101\cite{DBLP:journals/corr/ZagoruykoK16} model attained the lowest avg. precision, avg. recall, avg. F1-score and, test accuracy of 86.5\%, 85.5\%, 84.5\%, 85.0\% respectively.\\\\
    Among all the tested models, InceptionV3\cite{szegedy2015rethinking} and WideResNet101\cite{DBLP:journals/corr/ZagoruykoK16} attained the least accuracy of 88\% and 85\% respectively, while ViT-B/32\cite{dosovitskiy2021image} and DenseNet\cite{huang2018densely} were the best performers with accuracy of 97.6\% and 92\%. The ViT\cite{dosovitskiy2021image} model took around 10 minutes to train while other CNN based models took around 35 minutes. This proves that transformer based approaches consume less time when compared to CNN based approaches.
    Thus ViT\cite{dosovitskiy2021image} baseline model with 32x32 as input patch size is selected as the best model for classifying COVID-19 X-ray images.

\section{\normalsize{Conclusion and Future Work}}
    The COVID-19 viral infection is a global pandemic and is rapidly mutating. The lungs of the infected people are inflamed due to this viral infection. Hence, examining chest X-rays is one of the possible approaches in the detection of COVID-19. In this study, we have proposed an automated and accurate technique for distinguishing COVID-19 cases from Viral Pneumonia, Lung Opacity, and Healthy chest X-rays. We experimented with four different transfer learning-based architectures and, their performance is evaluated based on four performance metrics. The results of ViT-B/32\cite{dosovitskiy2021image} confirm that transformer-based models are on par with professional radiologists.\\\\
    Though this model achieves leading results in classifying COVID-19 chest X-rays, there is still scope for development. Noise is a key factor in radiography that affects the model's performance. Applying Generative Adversarial Network(GAN) based noise reduction \cite{Sun2018} techniques on the dataset can greatly improve the performance of our model. An ensemble learning-based approach can enhance the final performance. Using a large version of ViT\cite{dosovitskiy2021image} with a larger dataset can surely reach higher performance metrics. We are also planning to extend our work to the segmentation of COVID-19 chest X-rays \& CT scans to give even more insights for the radiologists.\\\\

\ifCLASSOPTIONcaptionsoff
  \newpage
\fi

\balance
\printbibliography
\end{document}